\documentclass{ws-procs9x6}

\usepackage{epsfig}

\begin{document}

\title{Nucleon and Hyperon Resonances with the Crystal Ball}

\author{W.J. Briscoe, A. Shafi, and I.I. Strakovsky}

\address{Department of Physics and Center for Nuclear Studies\\
The George Washington University, Washington, DC 20052, USA\\
E-mail: briscoe@gwu.edu}

\author{The Crystal Ball Collaboration}
\address{Abilene Christian University, Argonne National Laboratory,\\
Arizona State University, Brookhaven National Laboratory, \\
University of California at Los Angeles, University of Colorado,\\
George Washington University, Universit\"{a}t Karlsruhe,\\
Kent State University, University of Maryland, \\
Petersburg Nuclear Physics Institute, University of Regina,\\
Rudjer Boskovic Institute and Valparaiso University}

\maketitle

\abstracts{ The Crystal Ball Spectrometer is being used at
Brookhaven National Laboratory in a series of experiments which
study all neutral final states of $\pi ^{-}$p and $K^{-}$p induced
reactions. We report about the experimental setup and progress in
obtaining new results for the radiative capture reactions $\pi
^{-}p\rightarrow \gamma$n and $K^{-}p\rightarrow
\gamma\Lambda$,charge exchange $\pi^{-}p\rightarrow \pi^{\circ}$n,
two $\pi ^{\circ}$ production $\pi^{-}p\rightarrow \pi^{\circ}
\pi^{\circ}$n, and $\eta$ production $\pi^{-}p\rightarrow \eta n$
reactions. Data have also been obtained on the decays of $N^{*}$ ,
$\Delta $ , $\Lambda$ , and $\Sigma$ resonances. Threshold $\eta$
production has been studied in detail for both $\pi ^{-}$p and
$K^{-}$p. Sequential resonance decays have been investigated by
studying the 2$\pi^{o}$ production mechanism both in the
fundamental interaction and in nuclei. In addition, we have used
the $\eta$s produced near threshold to make precision measurements
searching in particular for rare and forbidden $\eta$ decays.}

\section{Introduction}

Quantum Chromodynamics, QCD, attempts to explain the strong
interaction in terms of the underlying quark and gluon degrees of
freedom. The study of the structure of baryons and their
excitations in terms of the elementary quark and gluon
constituents is thus pivotal to our understanding of nuclear
matter within QCD. It is the major goal of Nuclear Physics to
understand the strong interaction and keeping this goal in mind
motivates our choice of the particular reactions of interest that
we selected to study with the Crystal Ball at BNL. We will
describe these measurements below, but first we look at the
experimental conditions at BNL.

\section{The BNL Experimental Setup}

\subsection{The Crystal Ball}

We used the SLAC Crystal Ball in the C6 line at the Brookhaven
National Laboratory, BNL, Alternating Gradient Synchrotron, AGS,
with pion and kaon momenta up to 760 MeV/c. Data are taken
simultaneously on all reactions for each incident particle; this
helps ensure us that background events are accurately subtracted
and also allows us to normalize reactions with smaller cross
sections to more well determined reactions. Data taking using the
Crystal Ball began in July 1998 and continued until late November
1998. A second period of data taking occured in May 2002.

The Crystal Ball is a segmented, electromagnetic, calorimetric
spectrometer, covering 94\% of $4\pi $ steradians. It was built at
SLAC and used for meson spectroscopy measurements there for three
years. It was then used at DESY for five years of experiments and
put in storage at SLAC from 1987 until 1996 when it was moved to
BNL by our collaboration.

The Crystal Ball is constructed of 672 hygroscopic NaI crystals,
hermetically sealed inside two mechanically separate stainless
steel hemispheres. The crystals are viewed by photomultipliers,
PMTs. There is an entrance and exit tunnel for the beam, LH$_{2}$
target plumbing, and veto counters.

The crystal arrangement is based on the geometry of an icosahedron
(20 triangular faces or ``major-triangles'' arranged to form a
spherical shape.) Each ``major-triangle'' is subdivided into four
``minor-triangles'', which in turn consist of nine individual
crystals. Each crystal is shaped like a truncated triangular
pyramid, points towards the interaction point, is optically
isolated, and is viewed by a PMT which is separated from the
crystal by a glass window. The beam pipe is surrounded by 4
scintillators covering 98\% of the target tunnel (these
scintillators form the veto-barrel.)

This high degree of segmentation provides excellent resolution.
Electromagnetic showers in the Crystal Ball are measured with an
energy resolution of $\sigma/E = 2.7\%/E[GeV]^{1/4}$. Shower
directions are measured with a resolution in $\theta$ of $\sigma =
2^\circ$--$3^\circ$ for photon energies in the range 50--500 MeV;
the resolution in $\phi$ is $2^\circ/{ \sin \theta}$. Typically,
98\% of the deposited energy of each photon is contained in a
cluster of thirteen crystals (a crystal with its twelve nearest
neighbors). The thickness of the NaI amounts to nearly one hadron
interaction length resulting in two-thirds of the charged pions
interacting in the detector. The minimum ionization energy
deposited is 197 MeV; the length of the counters corresponds to
the stopping range of 233 MeV for $\mu^{\pm}$ , 240 MeV for
$\pi^{\pm}$, 341 MeV for $K^{\pm}$ , and 425 MeV for protons. The
preliminary energy calibration is performed using the 0.661 MeV
$\gamma$'s from a $^{137}C\!s$ source. The final energy
calibration is done using three reactions: i) $\pi
^{-}p\rightarrow \gamma n$\ at rest, yielding an isotropic,
monochromatic $\gamma $ flux of 129.4 MeV; ii) $\pi
^{-}p\rightarrow \pi ^{\circ }n$\ at rest, yielding a pair of
photons in the energy range 54.3---80 MeV, almost back to back;
and iii) $\pi ^{-}p\rightarrow \eta n$\ {\ }at threshold, yielding
two photons, about 300 MeV each, in coincidence almost back to
back. The PMT analog pulses are sent to ADCs for digitization.
Analog sums of the signals from each minor-triangle are available
for trigger purposes.

In addition to the expected high efficiency for photons, the
Crystal Ball is also fairly responsive to neutrons. We were able
to measure the response of the NaI(Tl) to neutrons by using the
reaction $\pi ^{-}p\rightarrow \pi ^{\circ }n$ and kinematics to
determine the efficiency (as high as 40\%) as a function of
energy.\cite{SS01}

\subsection{Beam Line}

\begin{figure}[htbp]
\centering{\
\psfig{file=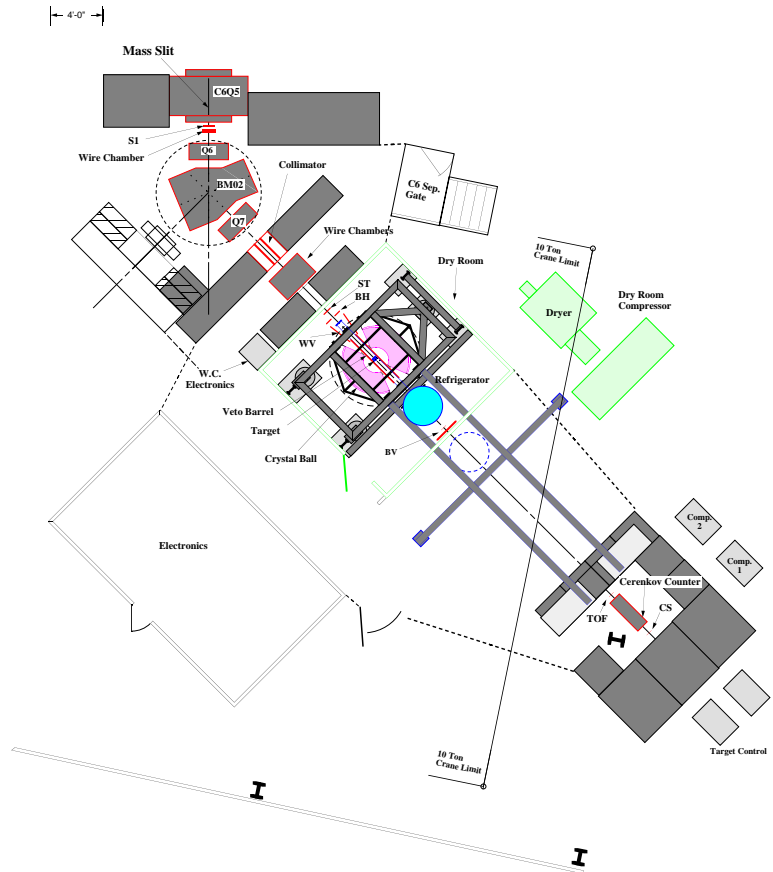,height=0.7\textheight,angle=0}}
\caption{ Experimental setup at the BNL AGS C6 line. \label{fig:r1r2}}
\end{figure}

Figure \ref{fig:r1r2} shows the experimental setup at BNL on the
C6 beam line. The final stages of the C6 beamline consist of four
quadrupoles and a dipole that form a beam momentum spectrometer.
Wire chambers are located on both sides of the dipole to track the
particles through the spectrometer. The momentum resolution is
0.3\%. The scintillators located up and down stream of the dipole
provide TOF information and the coincidence trigger for the beam.
Scintillators surround the LH2 target to provide a charged
particle veto. Two columns of scintillator neutron counters are
located downstream of the Crystal Ball. A beam veto scintillator
is located further downstream. A concrete shield wall located
upstream of the beam stop shields the Crystal Ball from low energy
photons from the stop. A \v{C}erenkov counter is located just
after this wall to monitor electron contamination in the beam.

The usual trigger consists of: a beam coincidence trigger, no
downstream beam veto, and a total energy-over-threshold signal
from the Crystal Ball. The trigger is normally set on the total
energy level. A trigger based on the distribution of the energy in
different regions of the Crystal Ball was also used to provide a
more restrictive trigger in certain cases.

\section{Pion-Induced Reactions}

\subsection{Radiative Capture}

The radiative decay of a resonance provides the ideal laboratory
for testing theories of the strong interaction, gives us insight
into the fundamental interactions between mesons and nucleons, and
allows us to probe into the structure of the nucleon itself. In
particular, these data are important in the study of the radiative
decay of the neutral Roper resonance. They can be combined with
recent JLab Hall B data for the reactions $\gamma p \rightarrow
\pi ^{+}$n and $\gamma p \rightarrow \pi ^{\circ}$p which study
the mesonic decays of the charged Roper resonance in the incident
photon energy region from 400 to 700 MeV. In addition, comparison
of our data to the new JLab data taken on the inverse reaction
$\gamma n \rightarrow \pi ^{-}$p, using a deuteron
target,\cite{BR94} tests extrapolation techniques for the deuteron
correction and study medium effects within the deuteron.

\begin{figure}[htbp]
\centering{\ 
\psfig{file=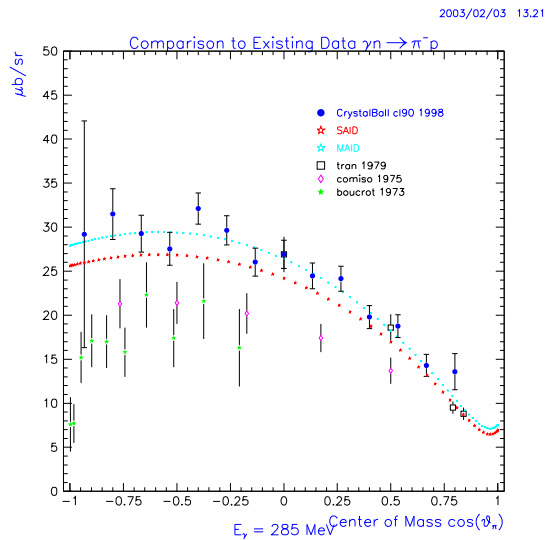,height=0.4\textheight,angle=0}}
\caption{ The radiative capture differential cross section
(plotted in the photoproduction channel) at a photon energy of 285
MeV - a pion beam momentum of 238 MeV/c.  The normalization factor
was obtained from the CEX - the effective beam flux was identical
for both these measurements (the same data run was analyzed for
the charge exchange as well as the radiative capture channels). In
this plot we also include previous data,\protect\cite{TR79}$^,$
\protect\cite{CO75}$^,$ \protect\cite{BO73} and the SAID
\protect\cite{AR02} as well as MAID \protect\cite{KA02}
predictions. \label{fig:rex}}
\end{figure}

Figure \ref{fig:rex} shows the radiative capture differential
cross section (plotted in the photoproduction channel) at a photon
energy of 285 MeV which corresponds to a pion beam momentum of 238
MeV/c.  The differential cross section was converted to the
photoproduction channel by using detailed balance; the
normalization factor was obtained from the first plot, since the
effective beam flux was identical for both these measurements (the
same data run was analyzed for the charge exchange as well as the
radiative capture channels).  In this plot, we also include
existing data,\cite{TR79}$^,$ \cite{CO75}$^,$ \cite{BO73} the
SAID\cite{AR02} prediction as well as the MAID\cite{KA02}
prediction.

\subsection{Charge Exchange}

The elusive charge exchange process has been the weakest link in
partial-wave and coupled-channel studies. The accurate data that
we obtained in this momentum region will help in improving the
determinations of the isospin-odd s-wave scattering length, the
$\pi$NN coupling constant, and the $\pi$-N $\sigma$ term. In
addition, better charge exchange data helps in evaluating the mass
splitting of the $\Delta$ - charge splitting of the P$_{33}$
resonance and may result in new values for the P$_{11}$(1440) mass
and width.

\begin{figure}[htbp]
\centering{\
\psfig{file=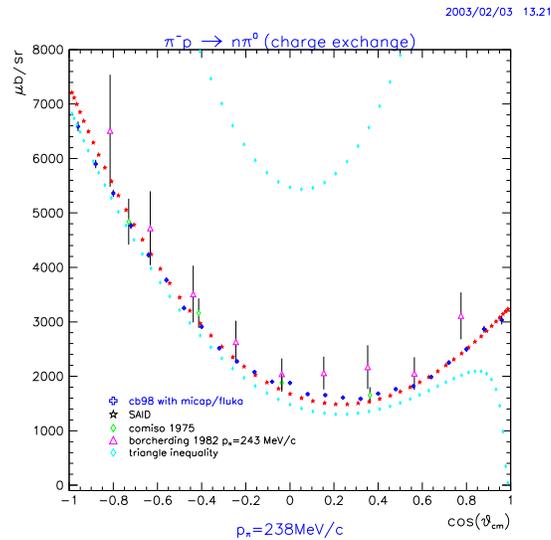,height=0.4\textheight,angle=0}}
\caption{Our charge exchange differential cross section compared
to previous measurements, \protect\cite{CO75}$^,$
\protect\cite{BO82} and the SAID \protect\cite{PA02} prediction at
a pion beam momentum of 238 MeV/c. The triangle inequality sets
upper and lower limits on the strength of charge exchange and is
based on $\pi^- p$ and $\pi^+ p$ elastic scattering data obtained
from SAID. \label{fig:cex}}
\end{figure}

Figure \ref{fig:cex} shows the measured charge exchange
differential cross section compared with the previously existing
data, and the SAID \cite{PA02} prediction at a pion beam momentum
of 238 MeV/c. We also include the triangle inequality which sets
upper and lower boundaries on the strength of charge exchange
based on $\pi^- p$ and $\pi^+ p$ elastic scattering data obtained
from SAID.~\cite{PA02}

\subsection{Two-Pion Production}

Two pion production provides a means of studying sequential pion
resonant decays; with neutral pions we study the $\pi \pi$
interaction in the absence of final-state Coulomb effects and
owing to isospin considerations there is no contribution of $\rho$
decay. The study of this process on the proton ($ \pi^-p
\rightarrow \pi^0 \pi^0 n$) is the subject of a recently completed
Ph.D. thesis. \cite{KC01} We have also published measurements on
this process in the nuclear medium.~\cite{AS00}

\subsection{Eta Production}

Near threshold $\eta$-production measurements provides
data\cite{KO02} useful in verifying models of $\eta$-meson
production and are also necessary for extraction of the $\eta$-N
scattering length. Precise $\eta$ production data are necessary to
resolve ambiguities in the resonance properties of the
S$_{11}$(1535) and in the $\eta$ photoproduction helicity
amplitudes.

\section{Kaon-Induced Reactions}

\begin{figure}[htbp]
\centering{\
\psfig{file=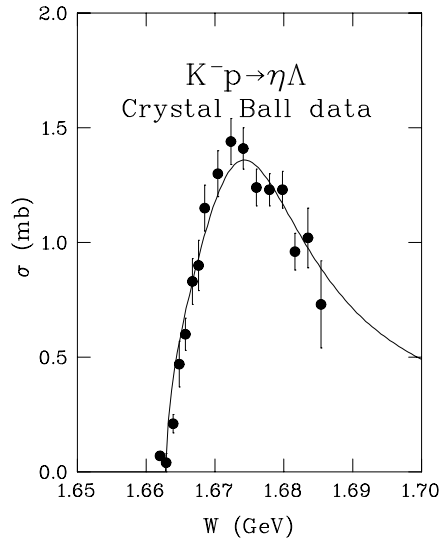,height=0.4\textheight,angle=0}}
\caption{Total cross section for the reaction $K^-p \rightarrow
\eta \Lambda$ . Solid squares show cross section derived from
$\eta \rightarrow \gamma \gamma$ and the open squares are derived
from the 3$\pi^0$ decay mode. \label{fig:el}}
\end{figure}

Using the Crystal Ball Detector, we have the ability of selecting
pure isospin states. For example in the reactions $K^-p
\rightarrow \eta \Lambda$ \cite{AS01} and $K^-p \rightarrow \pi^0
\Sigma^0$, we select a pure $I = 0$ $ \space$ $\Lambda^*$ and in
the reaction $K^-p \rightarrow \pi^0 \Lambda$, we select a $I = 1$
$\space$ $\Sigma^*$. Figure \ref{fig:el} shows the production
cross section of the former reaction. A significant part of our
program is geared toward the study of $K^-$p reactions are
described in a recent publication. \cite{MM01a}

\section{Decays of the Eta}

The large production cross section of tagged $\eta$s allow us to
search for breaking of fundamental symmetries ({\it e.g.} C and CP
invariance), and test Chiral Perturbation Theory as well as other
theoretical models. We have published an article on $\eta
\rightarrow$ 4$\pi^{o}$ \cite{SP00} which presents a new upper
limit for this branching ratio $(B \leq 6.9\times 10^{-7})$ of the
CP forbidden decay at the 90\% confidence level. This value of $B$
puts a 2\% limit on CP in quark-family-conserving interactions.

Another article has been published on the rare $\eta \rightarrow$
3$\pi^{o}$ \cite{BT01} which presents our determination of the
quadratic slope parameter $\alpha$ for that decay. The value
obtained $(\alpha = -0.031 \pm 0.004)$ disagrees significantly
with current theory. Since this published material is now readily
available, I will discuss our recent and yet unpublished work on
the $\pi^0 \gamma \gamma$ decay of the $\eta$.

\section{Rare Decays of the $\eta$ Meson}

As we have alluded to above, by studying the rare and forbidden
decay modes of the $\eta$ meson, we are able to test the limits of
such fundamental symmetries as C and CP invariance and G parity.
Additionally, these also provide a laboratory in which we can test
chiral perturbation theory and other recently proposed models. As
mentioned above, some of these measurements are already in the
literature.\cite{SP00} \cite{BT01}

To measure the $\eta$-decay processes, we took a series of dedicated $\pi^-$%
p runs at 720~MeV/c which was at the maximum in eta production and
yet close enough to threshold that the $\eta$s were essential
going forward in the lab. In effect, we produced an $\eta$ beam
and use it to measure the eta decay channels.

For the decay $\pi^- p \rightarrow \eta n \rightarrow \pi^0 \gamma
\gamma n$, we looked at events in which 4 photons were detected.
Even with this restriction, we still had backgrounds due to
3$\pi^0$ decay and direct 2$\pi^0$ production. These backgrounds
were reduced by a series of kinematic checks which not only
required that the 4 photons satisfied the kinematics of the
desired final state, but also eliminated any events with even a
small probability ($ 0.1\% $) of satisfying the kinematics of
possible background processes.

In addition to the above kinematic restrictions, we made various
target and detector cuts that tested our abilities to Monte Carlo
the acceptance and detection efficiencies. In all cases, we
obtained results consistent within our statistical and estimated
systematic uncertainties.

Our preliminary result for the $\eta \rightarrow \pi^0 \gamma
\gamma$ decay branching ratio is $3.2 \pm 0.9_{tot} \times
10^{-4}$.\cite{SP01} This is less than half of the current
Particle Data Group value of $7.1 \pm 1.4 \times 10^{-4}$ and
disagrees by about 3-4 standard deviations. However, our
experimental value does agree with the latest chiral perturbation
theory calculations.

Using a similar analysis procedure, one of our colleagues has just
reported an upper limit of the $\eta \rightarrow \pi^0 \pi^0
\gamma$ branching ratio of $5 \times 10^{-4}$ at the 90\%
confidence level.\cite{SP01a}

\section{Summary}

The Crystal Ball Program at BNL has producing a large number of
high-quality results in a very short time period after data
taking. We have put severe constraints on tests of chiral
perturbation theory and the limitations of fundamental symmetries.
Full and short reports as well as downloads of publications and
conference contributions are available to the public on our
Crystal Ball Collaboration web site.\cite{URL} While hoping to
complete our planned experimental program at BNL, members of the
collaboration have brought the Crystal Ball back to Europe, in
particular to MAMI at Mainz, Germany where a exciting new program
is about to begin.

\section{The Collaboration}

The new Crystal Ball Collaboration consists of B. Draper, S.
Hayden, J. Huddleston, D. Isenhower, C. Robinson, and M.
Sadler,{\em Abilene Christian University}, C. Allgower and H.
Spinka, {\em Argonne National Laboratory}, J. Comfort, K. Craig,
and A. Ramirez, {\em Arizona State University}, T. Kycia
(deceased), {\em Brookhaven National Laboratory}, M. Clajus, A.
Marusic, S. McDonald, B.~M.~K. Nefkens, N. Phaisangittisakul, and
W.~B.~Tippens, {\em University of California at Los Angeles}, J.
Peterson,
{\em University of Colorado}, W. Briscoe, A. Shafi, and I. Strakovsky {\em %
George Washington University}, H. Staudenmaier, {\em
Universit\"{a}t Karlsruhe}, D.~M.~Manley and J. Olmsted, {\em Kent
State University}, D. Peaslee, {\em University of Maryland}, V.
Abaev, V. Bekrenev, N. Kozlenko, S. Kruglov, A. Kulbardis, I.
Lopatin, and A. Starostin, {\em Petersburg
Nuclear Physics Institute}, N. Knecht, G. Lolos, and Z. Papandreou, {\em %
University of Regina}, I. Supek, {\em Rudjer Boskovic Institute}
and A.
Gibson, D. Grosnick, D.~D.~Koetke, R. Manweiler, and S. Stanislaus, {\em %
Valparaiso University}.

\section{Acknowledgements}

The members of the Crystal Ball Collaboration are supported in
part by the United States Department of Energy, the United States
National Science Foundation, the National Sciences and Engineering
Research Council of Canada, the Russian Ministry of Sciences,
Volkswagen Stiftung and the George Washington University Research
Enhancement Fund and Virginia Campus.

\end{document}